\documentclass{nature}
\usepackage{amsfonts}
\usepackage{mathrsfs,amssymb}
\usepackage[top=1in, bottom=1in, left=1in, right=1in]{geometry}
\usepackage{amsmath}
\usepackage{epsfig}
\usepackage{bm}
\usepackage{graphicx}
\usepackage{grffile}
\usepackage{ulem}
\usepackage{cancel}
\usepackage{cases}
\usepackage{mathtools}

\title{Nuclear quantum memory and time sequencing of a single $\gamma$ photon}

\author{Xiwen Zhang$^{1}$, Wen-Te Liao$^{1,2}$, Alexey Kalachev$^{3,4}$, Rustem Shakhmuratov$^{3,4}$, Marlan Scully$^{1,5}$ \& Olga Kocharovskaya$^1$}

\begin{document}

\maketitle

\begin{affiliations}
 \item Department of Physics and Astronomy, Texas A\&M University, College Station, Texas 77843, USA
 \item Department of Physics, National Central University, Taoyuan City 32001, Taiwan
 \item Zavoisky Physical-Technical Institute, Sibirsky Trakt 10/7, Kazan 420029, Russia
 \item Kazan Federal University, Kremlevskaya 18, Kazan 420008, Russia
 \item Baylor University, Waco, Texas 76706, USA
\end{affiliations}

\begin{abstract}
A $\gamma$-ray-nuclear quantum interface is suggested as a new platform for quantum information processing, motivated by remarkable progresses in $\gamma$-ray quantum optics. The main advantages of a $\gamma$ photon over an optical photon lie in its almost perfect detectability and much tighter, potentially sub-angstrom, focusability. Nuclear ensembles hold important advantages over atomic ensembles in a unique combination of high nuclear density in bulk solids with narrow, lifetime-broadening M\"ossbauer transitions even at room temperature. This may lead to the densest long-lived quantum memories and the smallest size photon processors. Here we propose a technique for $\gamma$ photon quantum memory through a Doppler frequency comb, produced by a set of resonantly absorbing nuclear targets that move with different velocities. It provides a reliable storage, an on-demand generation, and a time sequencing of a single $\gamma$ photon. This scheme presents the first $\gamma$-photon-nuclear-ensemble interface opening a new direction of research in quantum information science.
\end{abstract}

In the last decade optical-atomic interfaces have been developed as one of the basic building blocks for quantum information processing~\cite{Hammerer10}. However, optical photons are subjected to some practical and fundamental limitations, such as a lack of reliable inexpensive single-photon sources, a low efficiency and a high dark-count rate of single-photon detectors, and a $\sim 1$~$\mu$m diffraction limit imposing onto the size of information processing devices. These problems can be resolved in the $\gamma$-ray range, where i) single-photon detectors have nearly $100\%$ efficiency with almost no false detection, ii) radioactive decay in a cascade scheme produces heralded single $\gamma$ photons, and iii) sub-angstrom wavelength does not impose any practical limit on the size of a photonic circuit.

As for the atomic media, optical transitions are typically strongly broadened at room temperature, and narrow linewidths can be achieved only in a cryogenic environment at low atomic density. The M\"ossbauer nuclear transitions in bulk solids offer a solution to this problem. Even at room temperature and $\sim 10^{23}$~cm$^{-3}$ nuclear density, they may have narrow lifetime-broadening linewidths (in the range $1$~Hz - $1$~MHz), resulting in a very strong coupling of a single $\gamma$ photon with nuclear ensemble. For instance, just a $100$~nm-long stainless-steel film, $98\%$ enriched with $^{57}$Fe, would be optically thick for $\gamma$ photons resonant to the $14.4$~keV M\"ossbauer nuclear transition of $^{57}$Fe at room temperature. Therefore, such nuclear ensembles hold a good promise for the densest long-lived room-temperature quantum memories.

The tool box for coherent control of $\gamma$-ray-nuclear interaction has been immensely advanced in the past few decades, including recent development of the coherent sources in $10$~keV range, $\gamma$-ray mirrors, waveguides, cavities and beam splitters~\cite{Helisto91, Shvydko93, Shvydko96, Coussement02, Palffy09, Shvydko11, Amann12, Rohlsberger12, Shwartz12, Zhang13gamma, Adams13, Heeg13, Osaka13, Vagizov14, Shakhmuratov15, Heeg15, Liao15}, etc. Recently, the ability to shape a single $\gamma$ photon by vibrating nuclear resonant absorber was experimentally demonstrated, which allows one to encode information into a time-bin qubit~\cite{Vagizov14} or more complicated temporal waveforms of a single $\gamma$ photon~\cite{Shakhmuratov15}.

Quantum memory, representing itself as a controllable delay line for a single photon, lies in the heart of quantum computation and communication devices~\cite{Lvovsky09, Afzelius15}. Various techniques of quantum optical memory have been developed recently~\cite{Lvovsky09, Afzelius15,Chaneliere05,Reim11,Afzelius09,Hetet08}. However, a direct transfer of these techniques from optical to $\gamma$-ray range is hardly possible. Such optical techniques, as electromagnetically induced transparency~\cite{Chaneliere05}, off-resonant Raman~\cite{Reim11}, and atomic frequency comb (AFC)~\cite{Afzelius09}, imply a presence of strong coherent driving fields, which are not available yet. Meanwhile, as far as a gradient echo memory (GEM) technique~\cite{Hetet08} is concerned, it would require a strong (due to the small value of nuclear magneton) and fast-switchable external magnetic field (tens of tesla switched within $\sim 1$ nanosecond in the case of $^{57}$Fe~\cite{Adams11}).

\begin{figure}[Fig. 1]
\begin{center}
\epsfig{figure=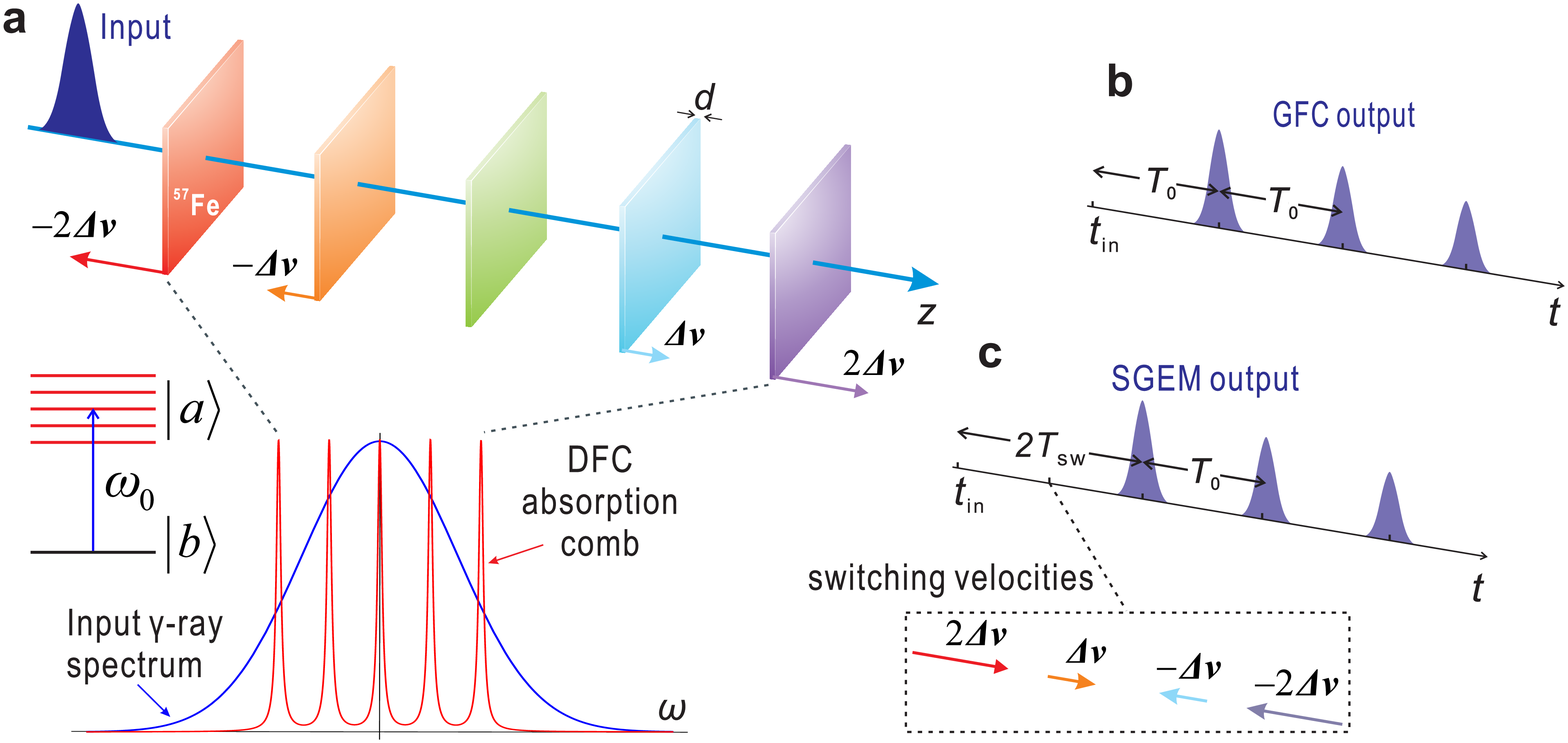, width=12cm}
\end{center}
\caption{\textbf{Illustration of $\gamma$-ray nuclear quantum memory.} \textbf{a,} The input single $\gamma$-photon wave packet is absorbed by the Doppler frequency comb, formed by a number of M\"{o}ssbauer targets, moving with velocities $v_m = m\Delta v$, $m=0,\pm 1, \pm 2, \cdots, \pm (M-1)/2 $. \textbf{b,} In GFC regime, the periodic beating between spectral components of the polarization in different targets forms the echo signal. \textbf{c,} In SGEM regime, the velocity directions of all targets are switched to the opposite at the moment $t_\text{in}+T_{\text{sw}}$ before the appearance of the first GFC echo. The phases of the targets' polarizations spread before switching and rewind after switching, so that an echo emerges at twice of the switching time $t_\text{in}+2T_{\text{sw}}$.}
\label{FigureScheme}
\end{figure}

We propose to store a single $\gamma$-ray wave packet of central frequency $\omega_0$, full-width-half-maximum (FWHM) field duration $\Delta t$ and arrival time $t_\text{in}$ in a two-level nuclear resonant medium composed of $M$ identical M\"ossbauer targets, which have the same optical thickness $\zeta^0$ and move at different velocities $v_m$ with equal velocity spacing $\Delta v$ (Fig. 1a). Due to Doppler effect, such velocity distribution forms a frequency comb in the resonant absorption spectrum of that set of targets with teeth separation $\beta \omega_0$, teeth width $2\Gamma$ (where $\beta = \Delta v /c$ and $\Gamma$ is a decay rate of nuclear coherence), and corresponding finesse $\mathcal{F}=\beta \omega_{0}/(2\Gamma )$. We call it a Doppler frequency comb.

A quantum storage with the Doppler frequency comb can be implemented in two different regimes. The first regime is similar to AFC~\cite{Afzelius09} quantum memory, except here the comb teeth are not present all together in each point as in AFC, but distributed along the photon propagation direction. Hence we name it a gradient frequency comb (GFC). As shown in Fig. 1b, because of beating between the discrete frequency field components re-emitted by different targets, echoes of the single $\gamma$ photon emerge at integer multiples of the rephasing time $T_0=2\pi /(\beta \omega _{0})$. In this work, unless otherwise specified, we mainly focus on the first echo pulse, which reads in GFC regime as (see Methods)
\begin{align}
\mathcal{E}_{\text{out}}(t)& \approx e^{-\frac{\pi }{4}\zeta_{\text{eff}}^{0}}\mathcal{E}_{\text{in}}(t)-
\frac{\pi \zeta_{\text{eff}}^{0}}{2}e^{-\frac{\pi \zeta_{\text{eff}}^{0}}{4}}e^{-\frac{\pi }{\mathcal{F}}}\mathcal{E}_{\text{in}}(t-T_{0}),
\label{slnI}
\end{align}
where $\mathcal{E}$ is the slowly varying amplitude of the $\gamma$-ray field and $\zeta^{0}_{\text{eff}}=\zeta^{0} /\mathcal{F}$ is an individual effective optical thickness. The first term of equation (\ref{slnI}) is the leakage field, i.e. the field not absorbed by the comb, and the second term represents the first GFC echo pulse. As an example, we consider an input single photon of duration $\Delta t=7$~ns with central frequency on resonance with the $14.4$~keV nuclear transition of $^{57}$Fe~\cite{Vagizov14, Shakhmuratov15}. The resonant medium consists of five $^{57}$Fe-enriched stainless-steel foils, each with $\zeta^0 = 41.3$ (corresponding to a total optical thickness $\zeta = M \zeta ^0 = 206.5$) and $\Delta v=3.075$~mm/s. As shown in Fig. 2a, the photon is retrieved with an efficiency $\approx 45\%$ after being stored for $28$~ns.

\begin{figure}[Fig. 2]
\begin{center}
\epsfig{figure=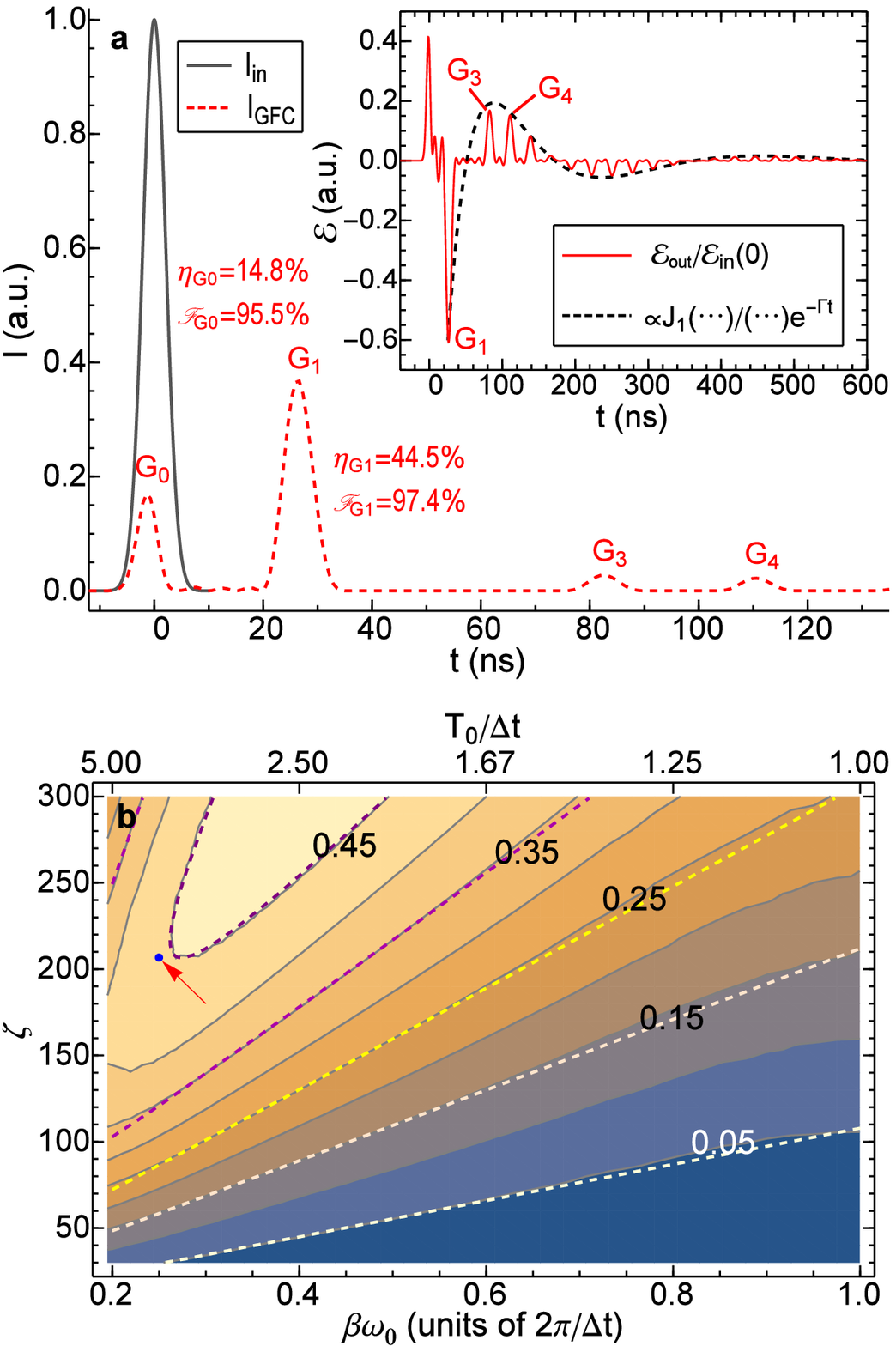, width=8.0cm}
\end{center}
\caption{\textbf{Gradient frequency comb (GFC) memory.} \textbf{a,} Input ($I_\text{in}$) and output ($I_\text{GFC}$) intensities as functions of time from numerical simulation of equations (\ref{EvolM}) and (\ref{EvolS}). The input field has FWHM duration $\Delta t = 7$~ns (intensity FWHM $4.95$~ns). The medium is composed of $M=5$ $^{57}$Fe-enriched M\"{o}ssbauer targets with $\Gamma / (2\pi) =0.55$~MHz, $\mathcal{F}=32.47$ and $\zeta_\text{eff}^0 = 1.27$. The peaks $G_{1,3,4}$ are GFC echoes ($T_{0}=28$~ns). The inset shows that higher sequence of GFC echoes are described by the response function $\propto e^{-\Gamma t} J_1(2\sqrt{\zeta^0 \Gamma t/2})/\sqrt{\zeta^0 \Gamma t /2}$ [see equation (S45) in supplementary information]. \textbf{b,} Numerical simulation (background contour plot) and analytical calculation [based on equation (\ref{slnI})] (dashed lines) of the GFC echo efficiency (see definition in Methods) as a function of the total optical thickness $\zeta = M \zeta^0$ and the frequency comb spacing $\beta \omega_0$ for the same parameters ($M$, $\Delta t$, $\Gamma$) as in \textbf{a}, satisfying the constrain given by equation (\ref{Condition2}). Graph \textbf{a} is plotted for the parameters corresponding to the point indicated by the arrow.} \label{GFCEfficiency7nsI2nomT}
\end{figure}

According to equation (\ref{slnI}), the upper bound of the first GFC echo efficiency is $54\%$, which can be achieved by optimizing the optical thickness and finesse at the following conditions:
\begin{align}
& \zeta^{0} \approx \frac{4}{\pi }\mathcal{F}\gg 1,  \label{Condition1} \\
& \frac{\pi }{M\Delta t \Gamma}<\mathcal{F}<\frac{\pi }{\Delta t \Gamma}. \label{Condition2}
\end{align}
From equation (\ref{Condition1}), in order to reduce the role of decoherence, a high finesse is required. But at high finesse the portion of the full comb bandwidth covered by the comb teeth is too small to retain the input energy. To achieve the optimal storage efficiency, one has to effectively broaden each comb tooth by means of optical thickness such that $\zeta^0 2\Gamma\approx \beta \omega_0$ in accordance with equation (\ref{Condition1}). Additional condition (\ref{Condition2}) is required to ensure spectrum coverage and echo's temporal resolvability, which is clear after being unfolded into $1/M < \beta \omega_0/(2\pi/\Delta t) <1$ or $ \Delta t < T_0 < M \Delta t$. The GFC regime can also be used as a way to split a single-peak $\gamma$ photon into a time-bin waveform. An equal splitting of an input photon between the leaked and delayed fractions of the output photon is achieved at the optical thickness $\zeta_{\text{eff}}^0 = (2/\pi)e^{\pi/\mathcal{F}}$, with conversion efficiency $50\%$ for $\mathcal{F} = 10$.

\begin{figure}[Fig. 3]
\begin{center}
\epsfig{figure=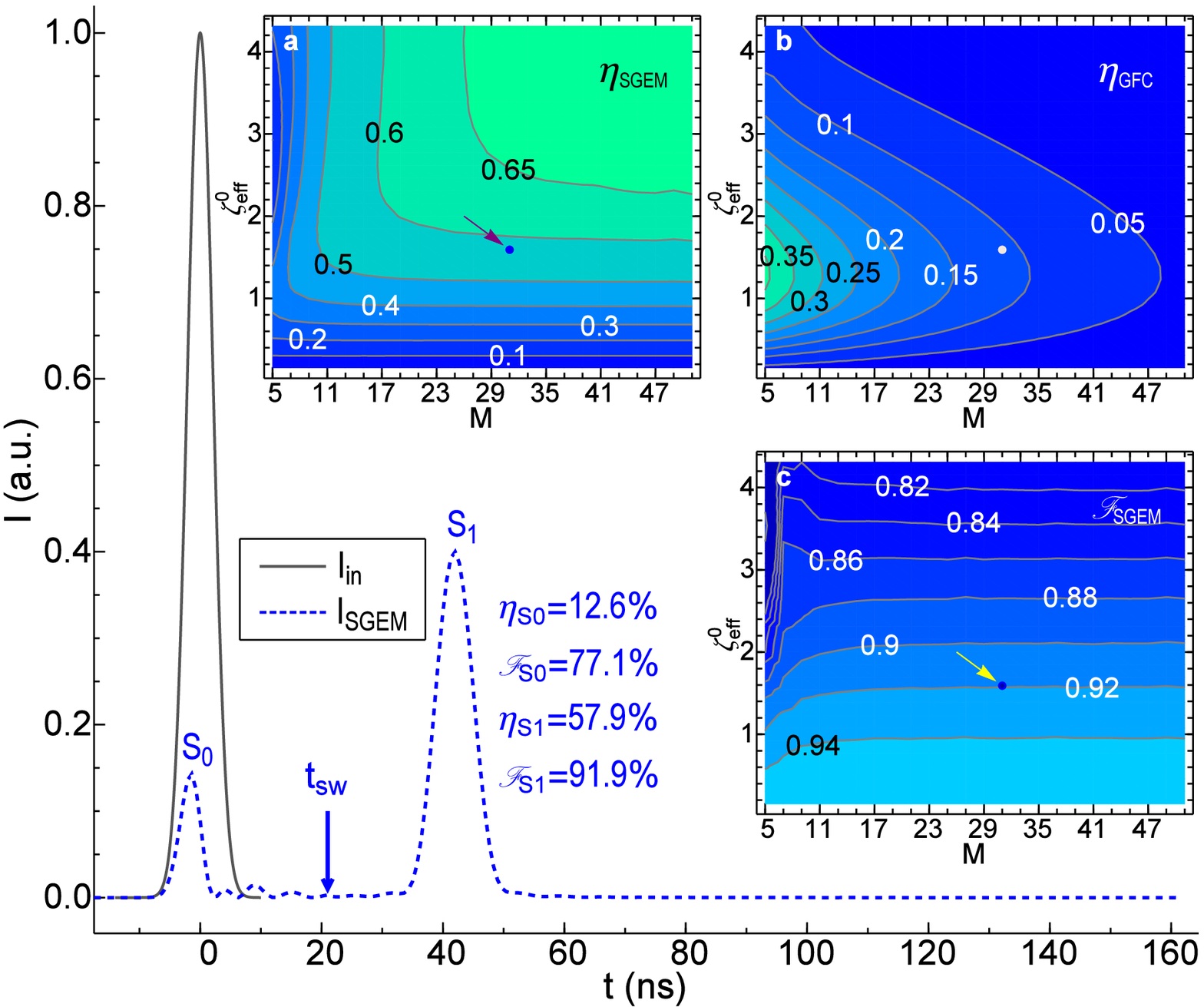, width=10.00cm}
\end{center}
\caption{\textbf{Stepwise gradient echo memory (SGEM).} Input ($I_\text{in}$) and output ($I_\text{SGEM}$) $\gamma$ photon intensities as functions of time for $M = 31$ targets, each with an optical thickness $\zeta^0 = 206.7/31=6.67$. Inset \textbf{a-b,} Numerical simulation of SGEM and GFC first echo efficiencies, respectively, as functions of individual effective optical thickness $\zeta^0_\text{eff}$ and number of targets $M$. Inset \textbf{c,} Numerical simulation of SGEM first echo fidelity $\mathscr{F}$ (see definition in Methods) as a function of $\zeta^0_\text{eff}$ and $M$. All figures are plotted for the pulse duration $\Delta t=7$~ns, switching time $T_\text{sw} = 21$~ns, and fixed comb bandwidth $M\beta\omega_0 = 2\pi / \Delta t$. The arrows in the insets indicate the parameters corresponding to the main plot.}\label{SGEMefficiency}
\end{figure}

Essentially higher efficiency than the upper bound of GFC echo $54\%$ can be achieved using another regime of Doppler frequency comb scheme, which we call a stepwise gradient echo memory (SGEM). It can be implemented by switching the directions of motion of all targets to the opposite ($v_m \to -v_m$) at some moment of time $t_\text{in} + T_{\text{sw}}$, before the appearance of the first GFC echo (see Fig. 1c). Such switch enforces a rewind of the phase evolution of the polarizations in the moving targets. This regime is similar to GEM~\cite{Hetet08}, except here the frequency of the resonant transition is changed not continuously but in a stepwise manner along the propagation direction. The echo is formed when the phases regress back to their original state. The switch time is chosen to satisfy $\Delta t/2 < T_{\text{sw}} < T_0 - \Delta t/2$, so that the SGEM echo appears as the first retrieval signal (see Fig. 3). Thus the storage time of the signal, $2T_\text{sw}$, can be completely controlled over the time interval $(\Delta t, 2T_0-\Delta t)$, allowing to produce a $\gamma$ photon on demand. In the limit of large number of targets ($M\to \infty$) SGEM regime transforms into GEM scheme, which may provide $100\%$ efficiency when the decoherence effects are small enough~\cite{Moiseev08}. Hence simply by splitting the same total optical thickness into more targets, we can increase the storage efficiency and, in particular, make it higher than the theoretical upper limit of the efficiency in the GFC regime (see Fig. 3).

\begin{figure}[Fig. 4]
\begin{center}
\epsfig{figure=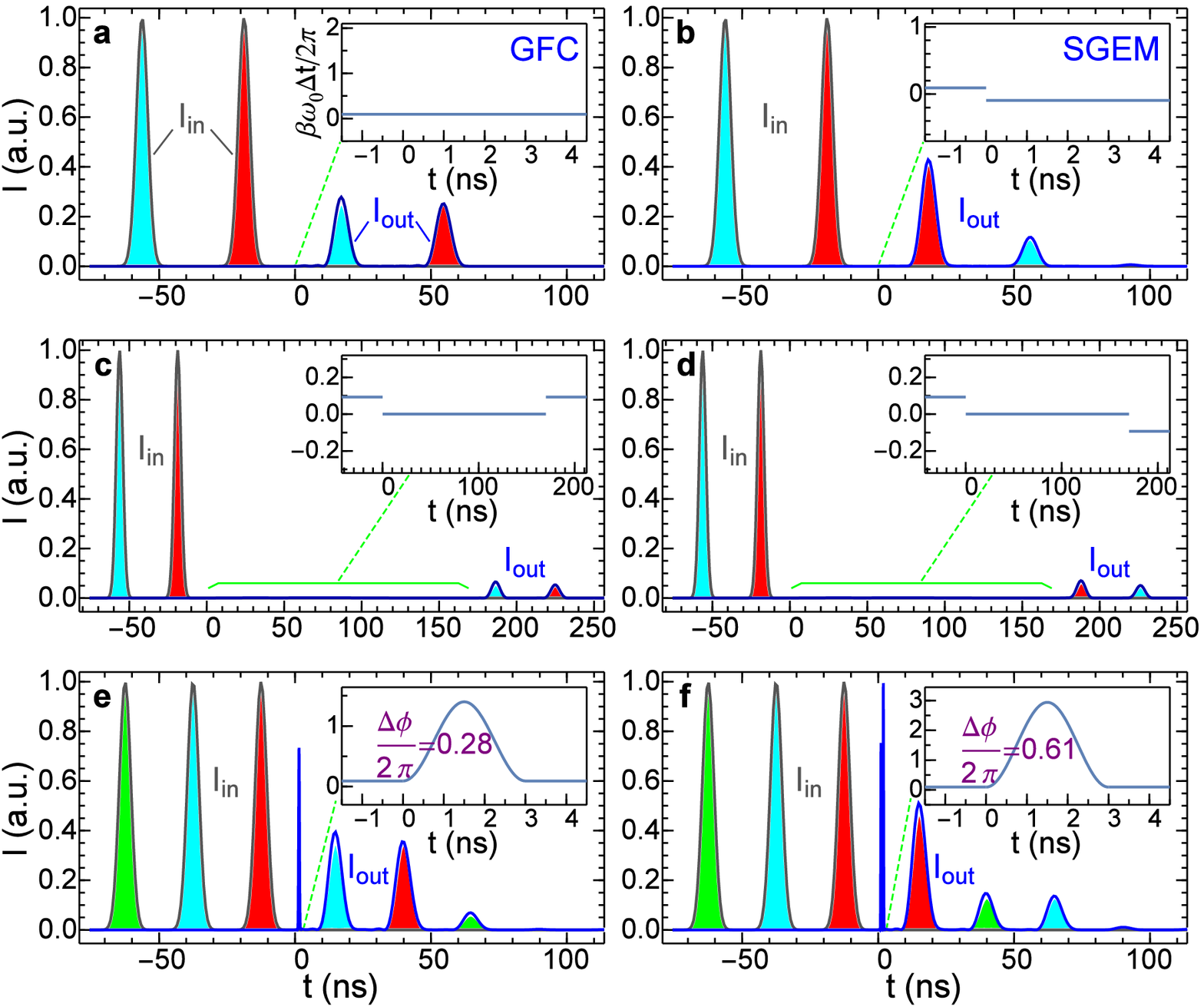, width=8.6cm}
\end{center}
\caption{\textbf{Single $\gamma$ photon processing.} The multi-peak input and output signals as functions of time (leakage at $t<0$~ns is not plotted). The filled colors correspond to single-peak input (output) signals. The insets show the velocity modulation. The common parameters are: $\Delta t = 7$~ns, $T_0$ = 75~ns, $\zeta^0 = 15.4$ and $M = 13$. \textbf{a-b,} GFC and SGEM echoes. \textbf{c-d,} Holding a double-peak signal by stopping all targets for $170$~ns (i.e. longer than excited state lifetime, 141 ns) and retrieving it in the same (\textbf{c}) and reversed (\textbf{d}) order. \textbf{e-f,} Circular permutations of the input triple-peak signal via sinusoidal modulation of the velocities that produces an additional phase difference $\Delta \phi / (2\pi)$ equal to $0.28$ (\textbf{e}) and $0.61$ (\textbf{f}). The narrow peaks in the vicinity of $t=0$ are the losses during the modulations.}\label{PhaseMod}
\end{figure}

The Doppler frequency comb allows one to realize not only storage, but also a variety of single-photon processing functionalities, including reversing of the photon's temporal shape, delayed and/or advanced retrieval, relative amplitude manipulation, temporal permutation, etc., which can be achieved by a modulation of the targets' velocities before the emergence of the echo. For example, by choosing either GEM or SGEM regime one can retrieve a time-bin qubit in the same or reversed order of the input signal (Fig. 4a-b). By stopping all targets after absorption, one can hold the echo for an arbitrary time (Fig. 4c-d). By boosting all targets' velocities via increasing velocity spacing $\Delta v$ to $\Delta v^\prime$ and back to $\Delta v$ in a time interval $t \in (t_i^\prime, t_f^\prime)$, one can impose additional phase difference $\Delta \phi = (\Delta v^\prime - \Delta v) \omega_0 (t_f^\prime - t_i^\prime)/c \in [0, 2\pi]$ between polarizations of two adjacent targets. Consequently, the first echo right after modulation will emerge at shifted moment of time $t_\text{in} +\big(p - \frac{\Delta \phi}{2\pi} \big) T_0$, where $p=\left\lceil \frac{t_f^\prime - t_\text{in}}{T_0}+\frac{\Delta \phi}{2\pi}\right\rceil$, and $\left\lceil x \right\rceil$ represents the smallest integer greater than or equal to $x$. In this way it becomes possible to manipulate the time of appearance of any individual peak from the incoming photon's waveform. For such processing, only the total phase difference matters, so that the modulation does not need to be square-shape (Fig. 4e-f).

In conclusion, we present the first $\gamma$-photon-nuclear-ensemble interface using a set of moving resonant nuclear targets for single $\gamma$ photon quantum memory and processing. Such quantum interface may become an extension of the existing optical and transportive technologies: the optical-electronic and $\gamma$-photon-nuclear evolutions are essentially decoupled and therefore can in principle coexist in one solid-state platform for integrated functionalities, opening up a new channel with a rich variety of interesting possibilities in the field.


\begin{methods}
Let the $m^{\text{th}}$ M\"{o}ssbauer target have the initial central position $l_m$, thickness $d$, nuclear density $N$, and resonant frequency detuning $\Delta_{m}= m \beta\omega_{0}$. The light-matter interaction in a one-dimensional model is described by the Maxwell-Bloch equations (see supplementary information for details):
\begin{align}
& \frac{\partial }{\partial z}\mathcal{E}(z,t)=\sum\limits_{m=-M_0}^{M_0} \mathcal{P}^{m}(z,t)\left( \Theta^{m}_{-} -\Theta^{m}_{+} \right),  \label{EvolM} \\
& \frac{\partial }{\partial t}\mathcal{P}^{m}(z,t)=(-\Gamma -i\Delta _{m}) \mathcal{P}^{m}(z,t)-|g|^2 N \mathcal{E}(z,t),
\label{EvolS}
\end{align}
where $\mathcal{P}^{m}$ is proportional to the slowly varying amplitude of the off-diagonal element of the density matrix for the $m^\text{th}$ target, $g$ is the coupling constant of the $\gamma$ ray-nucleus interaction, and $\Theta^{m}_{\pm}=\Theta( z-l_{m}\mp d/2)$ is the Heaviside step function.

In the GFC regime under conditions given by (\ref{Condition2}) and $\mathcal{F}\gg 1$, the solution of equations (\ref{EvolM}) and (\ref{EvolS}) yields the temporal Fourier transform of the output field as follows:
\begin{align}
& \mathcal{E}_{\text{out}}(\omega ) \approx  \notag \\
&\mathcal{E}_{\text{in}}(\omega)e^{-\frac{\pi }{4}\zeta_{\text{eff}}^{0}}\prod\limits_{n=1}^{\infty}\sum_{q=0}^{\infty }\bigg(-\frac{\pi }{2}\zeta_{\text{eff}}^{0}e^{-\frac{\pi n}{\mathcal{F}}}\bigg)^{q}\frac{e^{inq\omega T_{0}}}{q!}, \label{aoutexplicitinFreq}
\end{align}
where $\zeta^{0}_\text{eff}=2|g|^2 Nd/(\Gamma \mathcal{F})$. From equation (\ref{aoutexplicitinFreq}) we obtain equation (\ref{slnI}) that describes GFC signal up to $t=t_\text{in}+T_{0}$.

The action of the quantum memory scheme is characterized by the total efficiency $\eta$ and fidelity $\mathscr{F}$. For the first echo signal the former is defined as $\eta =N_{\text{out}} / N_{\text{in}}$, where $N_{\text{in}}=\int_{t_\text{in}-0.5t_\text{ec} }^{t_\text{in}+0.5t_\text{ec}}dt \mathcal{E}_{\text{in}}^{\dag}(t)\mathcal{E}_{\text{in}}(t) $ and $N_{\text{out}}=\int_{0.5t_\text{ec}}^{1.5t_\text{ec}}dt \mathcal{E}_{\text{out}}^{\dag }(t)\mathcal{E}_{\text{out}}(t) $, and the later is defined as
\begin{align}
\mathscr{F}=\frac { \left\vert\int_{t_\text{in}-0.5t_\text{ec} }^{t_\text{in}+0.5t_\text{ec}}dt \mathcal{E}_{\text{in}}^{\dagger }(t)\mathcal{E}_{\text{out}}\textbf{(}\pm (t- t_\text{in})+t_\text{ec} \textbf{)} \right\vert ^{2} } {N_{\text{in}}N_{\text{out}}}, \notag
\end{align}
in which ``$+$" is for GFC regime and ``$-$" is for SGEM regime, $t_\text{ec}$ is the appearance time of the (first) echo pulse, which is equal to $T_0$ for GFC and $2T_{\text{sw}}$ for SGEM. The GFC echo efficiency according to equation (\ref{slnI}) is:
\begin{align}
\eta _\text{G1}& =\left( \frac{\pi \zeta_{\text{eff}}^{0}}{2}e^{-\frac{\pi\zeta_{\text{eff}}^{0}}{4}}e^{-\frac{\pi }{\mathcal{F}}}\right) ^{2}.
\label{efficiency1}
\end{align}

The upper bound of the SGEM echo efficiency can be roughly estimated as follows. Let us assume that the ratio of the retrieved over the stored energy is the same as the ratio of the stored over an input energy, which is $\big(1-e^{-\pi \zeta_{\text{eff}}^{0}/2}\big)e^{-2\Gamma T_{\text{sw}}}$. Then the efficiency is on the order of $\sim \big(1-e^{-\pi \zeta_{\text{eff}}^{0}/2}\big)^2e^{-4\Gamma T_{\text{sw}}}$. In the limit of very large number of targets, corresponding to GEM, this is in agreement with the well-known analytical result~\cite{Moiseev08}. In particular, a storage of $7$~ns photon for $42$~ns in $^{57}$Fe with total optical thickness $206.7$ (see Fig. 3) is limited by $63\%$ efficiency. Higher efficiency could be achieved for storage of a shorter photon in a smaller interval of time during which the incoherent decay remains negligible. However, the storage of shorter pulses would imply larger bandwidth of the Doppler frequency comb, larger number of targets, and proportionally increased total optical thickness. Recently developed transducers~\cite{Sorokin13} provide modulation frequency up to $10$~GHz, which can be used both for production of short single $\gamma$-photon waveforms~\cite{Vagizov14, Shakhmuratov15} and for large comb bandwidths.
Longer photons with duration over a few nanoseconds can be efficiently stored in targets with longer lived M\"{o}ssbauer nuclear transitions, such as $93.3$~keV transition in $^{67}$Zn with coherence time $13.6$~$\mu$s~\cite{Shvydko90}.
M\"{o}ssbauer transitions with lifetimes much longer than tens of microseconds, such as $12.4$~keV transition in $^{45}$Sc with lifetime $0.46$~s and $88.0$~keV transition in $^{109}$Ag with lifetime $57.1$~s~\cite{Shvydko90}, are typically inhomogeneously broadened due to magnetic dipole-dipole interactions~\cite{Boolchand88}. Potentially, these interactions may be suppressed using techniques similar to those developed in nuclear magnetic resonance (see Ref. 30 and references therein), providing extraordinarily long storage time. Some of these elements demonstrate interesting optical-electronic properties as well. For example, silver is a well-studied surface plasmonic material for potential miniaturization of optical circuits, and zinc oxide is a rapidly developed semiconductor for optoelectronic devices.



\end{methods}




\begin{addendum}
 \item [Supplementary Information] is available in the online version of the paper.
 \item We gratefully acknowledge the support of the National Science Foundation (Grant No. PHY-$150$-$64$-$67$ and PHY-$130$-$73$-$46$). X.Z. is supported by the Herman F. Heep and Minnie Belle Heep Texas A$\&$M University Endowed Fund held/administered by the Texas A$\&$M Foundation. A.K. acknowledges the support from the Russian Science Foundation (Grant No. $14$-$12$-$00806$). R.S. acknowledges the support from the Program of Competitive Growth of Kazan Federal University funded by the Russian Government.
 \item[Author Contributions] X.Z., W.L., A.K. and O.K. developed the theoretical model and major equations. X.Z. derived the analytical solutions, did the numerical simulations, and wrote the paper with supplementary information. W.L. and A.K. independently performed simulations to check some of the results, and W.L. numerically verified full set of equations and slow-motion approximation used by X.Z. in the analytical derivation. A.K. and M.S. suggested important physical insights into the results of simulations. R.S. and O.K. proposed two mechanisms of Doppler frequency comb memory, based on stepwise gradient echo and gradient frequency comb. A.K. and X.Z. suggested and designed the single $\gamma$ photon sequencing. M.S. and O.K. coordinated the efforts. All authors discussed the results and edited the manuscript.
 \item[Competing Interests] The authors declare no competing financial interests.
 \item[Correspondence] Correspondence and requests for materials should be addressed to X.Z.~(xiwen@physics.tamu.edu or permanent email xiwen@mail.ustc.edu.cn) or to O.K.~(kochar@physics.tamu.edu).
\end{addendum}

\end{document}